\documentclass[twocolumn,prl,superscriptaddress,amsmath,amssymb]{revtex4}
\usepackage{graphicx,color,}

\begin{document}

\title{Front-mediated melting of isotropic ultrastable glasses}

\date{\today}

\author{Elijah Flenner}

\affiliation{Department of Chemistry, Colorado State University, Fort Collins CO 80523, USA}

\author{Ludovic Berthier}

\affiliation{Laboratoire Charles Coulomb (L2C), University of Montpellier, CNRS, Montpellier, France}

\author{Patrick Charbonneau}

\affiliation{Department of Chemistry, Duke University, Durham, NC 27708, USA}

\affiliation{Department of Physics, Duke University, Durham, NC 27708, USA}

\author{Christopher J. Fullerton}

\affiliation{Laboratoire Charles Coulomb (L2C), University of Montpellier, CNRS, Montpellier, France}

\affiliation{Department of Physiology, Anatomy and Genetics, University of Oxford, Oxford, United Kingdom}

\begin{abstract}
Ultrastable vapor-deposited glasses display uncommon material properties. Most remarkably, upon heating they are believed to melt via a liquid front that originates at the free surface and propagates over a mesoscopic crossover length, before crossing over to bulk melting. We combine swap Monte Carlo with molecular dynamics simulations to prepare and melt isotropic amorphous films of unprecedendtly high kinetic stability. We are able to directly observe both bulk and front melting, and the crossover between them. We measure the front velocity over a broad range of conditions, and a crossover length scale that grows to nearly $400$ particle diameters in the regime accessible to simulations. Our results disentangle the relative roles of kinetic stability and vapor deposition in the physical properties of stable glasses. 
\end{abstract}

\maketitle

Amorphous films created by physical vapor deposition have the same kinetic stability as liquid-cooled glass films aged for
thousands of years~\cite{Swallen2007}. Such exceptional stability makes these materials promising for a wide range of applications, including drug delivery~\cite{Hancock1997}, protective coatings~\cite{Yu2013,Chu2012}, and lithography~\cite{Neuber2014}. There is acute interest in better controlling the physical characteristics of these ultrastable glasses, and especially the way they lose their stability~\cite{Swallen2009,Kearns2010}. Ordinary liquid-cooled glasses melt homogeneously from the bulk, but ultrastable films are reported to melt via a constant-velocity front initiated at the free surface, a process reminiscent of the surface melting of crystals~\cite{Peng2010}. This analogy is however difficult to rationalize theoretically~\cite{Wolynes2009,Jack2016} because, contrary to crystal melting, glass and liquid are not distinct thermodynamic phases. The relative contribution of molecular structures and film anisotropy inherent to the nonequilibrium vapor deposition process, on the one hand, and of the kinetic stability on the melting kinetics, on the other hand, are not easily disentangled experimentally. 

Melting fronts in vapor-deposited glasses have been indirectly inferred in experiments~\cite{Rodriguez2015,Rafols2017,Rodriguez2014,Kearns2010,Swallen2010,Sepulveda2012,Chen2013,Sepulveda2013,Dalal2015,Tylinski2015,Walters2015,Sepulveda2014,marta-new}, observations using spectroscopic ellipsometry offering the most direct probe of the propagating front~\cite{Walters2015,Dalal2015}.
Generically, melting depends both on the initial preparation temperature of the film, $T_\mathrm{i}$, and on the temperature where melting is performed, $T_\mathrm{m}$. (Because the crystal plays no role in this process, $T_\mathrm{m}$ is unrelated to crystal melting.)
Experiments suggest that ultrastable glass films melt via a well-defined front that propagates from the free surface at constant velocity, $ v=v(T_\mathrm{i}, T_\mathrm{m})$, a phenomenon without equivalent in conventionally prepared glasses. Quantitatively, $v$ decreases when the glass stability (controlled by $T_\mathrm{i}$) increases and increases with $T_\mathrm{m}$. Two different functional forms were used to interpret the results: (i) an Arrhenius scaling~\cite{Walters2015,Tylinski2015}, $v =v_0 (T_\mathrm{i}) e^{-E_\mathrm{a}/T_\mathrm{m}}$, where $E_\mathrm{a}$ is an activation energy (with Boltzmann constant set to unity); and (ii) a power-law scaling
\begin{equation}
v =C(T_\mathrm{i}) \tau_\alpha(T_\mathrm{m}) ^{-\gamma},
\label{eq:alpha}
\end{equation} 
where $\gamma \leq 1$, and $\tau_\alpha(T_\mathrm{m})$ is the equilibrium bulk structural relaxation time at $T_\mathrm{m}$~\cite{Sepulveda2012,Rodriguez2015,Tylinski2015,Sepulveda2014}. The prefactor of both forms captures the stability dependence of $v$ encoded in $T_\mathrm{i}$. In thick films, the melting front propagates over a finite distance, $\ell_\mathrm{c} = \ell_\mathrm{c}(T_\mathrm{i},T_\mathrm{m})$, because deeper layers have homogeneously melted by a distinct bulk-like mechanism by the time the front reaches them. 
Available data suggest that $\ell_\mathrm{c}$ can vary from 20 to 2000 times the molecular size~\cite{Sepulveda2014,Sepulveda2012}, smaller $\ell_\mathrm{c}$ being reported for less stable systems and higher $T_\mathrm{m}$~\cite{Rafols2017,marta-new}. Such a large length scale characterizing the dynamics of supercooled liquids is surprising in materials that are structurally homogeneous down to the molecular size. Its physical origin and theoretical interpretation remain unclear, and understanding its evolution with physical parameters would enlighten its interpretation.

At the theoretical level, a treatment based on an extended mode-coupling theory predicts an interplay between front-mediated and bulk melting~\cite{Wolynes2009}, with the melting front being triggered by the increased mobility of top layers. Dynamical facilitation can also be analyzed using kinetically constrained models to shed light on front melting~\cite{Gutierrez2016,Leonard2010}. A three-dimensional East model suggests that the Arrhenius scaling of the front velocity breaks down at low melting temperatures, and that Eq.~\eqref{eq:alpha} with $\gamma=0.95$ then holds for all $T_\mathrm{m}$~\cite{Gutierrez2016}. A modified East model predicts the existence of a characteristic film thickness $\ell_\mathrm{c}$. Similar conclusions were drawn from another constrained model, reporting $\gamma \approx 0.83$~\cite{Leonard2010}. A two-dimensional plaquette model illustrates a nucleation and growth picture of melting for sufficiently stable glasses~\cite{Jack2016}, and evinced that the large associated length scale is related to $\ell_\mathrm{c}$. Numerical simulations could be expected to critically assess these various proposals, but limited attempts have been made because preparing glassy films sufficiently large and stable for a melting front to develop has been computationally too challenging.  Direct simulation of the vapor deposition process does not provide a sufficiently large gain in stability~\cite{Flenner2017,Lyubimov2013}, and the resulting evidence for front melting is incomplete~\cite{Lyubimov2013}. Although random pinning~\cite{Hocky2014} can create fairly stable inhomogeneous two-dimensional films, a surpringly modest $\ell_\mathrm{c}$ growth is observed, associated with $\gamma \approx 1$. Clearly, a generic predictive picture has yet to emerge from these theoretical studies.  

Here, we exploit a recent major advance in sampling methods for glass-forming liquids~\cite{Berthier2016,Ninarello2017} to create isotropic, three-dimensional films of exceptional thickness, homogeneity and kinetic stability for a model glass former. 
These features allow us to observe and quantitatively characterize the emergence of front-mediated melting and its competition with the bulk process, and hence to distinguish between glass stability and film preparation in controlling the melting kinetics.   This allows us to resolve the applicability of Eq.~\eqref{eq:alpha} and provide a robust estimate of the characteristic length scale $\ell_\mathrm{c}$. Our results show that kinetic stability and melting temperature are the key control parameters in the front-melting process of stable glasses, and that very large $\ell_\mathrm{c}$ emerge naturally in the non-equilibrium melting  of ultrastable glasses, even when starting from equilibrium isotropic films.    

We consider a glass-former composed of size polydisperse Lennard-Jones particles with pair interaction $V_{nm}(r) = \epsilon [(\sigma_{nm}/r)^{12} - (\sigma_{nm}/r)^6]$, where each particle $n$ has
a diameter $\sigma_n$ randomly chosen from the distribution $P(\sigma) = A/\sigma^3$, with $\sigma \in[0.73,1.62]$ and normalization constant $A$. We choose a nonadditive mixing rule, $\sigma_{nm} = \frac{1}{2}(\sigma_n + \sigma_m)(1-\Delta |\sigma_n - \sigma_m|)$
with $\Delta = 0.2$, in order to suppress fractionation and ordering at low pressure, $P$, and temperature, $T$. The unit of energy is set to $\epsilon$, the unit of length is the average particle diameter, $\sigma_0$, and, because all particles have the same mass $m$, the 
unit of time is $\sqrt{\sigma_0^2 m/\epsilon}$. 
To characterize bulk systems, we first thermalize configurations with $N=2000$ particles from constant $NPT$ swap Monte Carlo simulations at $P=0$~\cite{Ninarello2017}. Dynamical properties are then determined using standard $NVT$ Nos\'e-Hoover molecular dynamics simulations
with the Large-scale Atomic/Molecular Massively Parallel Simulator (LAMMPS)~\cite{lammps}.
We measure the time decay of the self-intermediate scattering function
$F_s(q;t) = N^{-1} \left< \sum_{n} e^{-i \mathbf{q} \cdot [\mathbf{r}_n(t) - \mathbf{r}_n(0)]} \right>$, where $q=7.1$  maximizes the static structure factor, and define the structural relaxation time, $\tau_\alpha$, as $F_s(q;\tau_\alpha) = e^{-1}$. 

\begin{figure}
\includegraphics[width=1.\columnwidth]{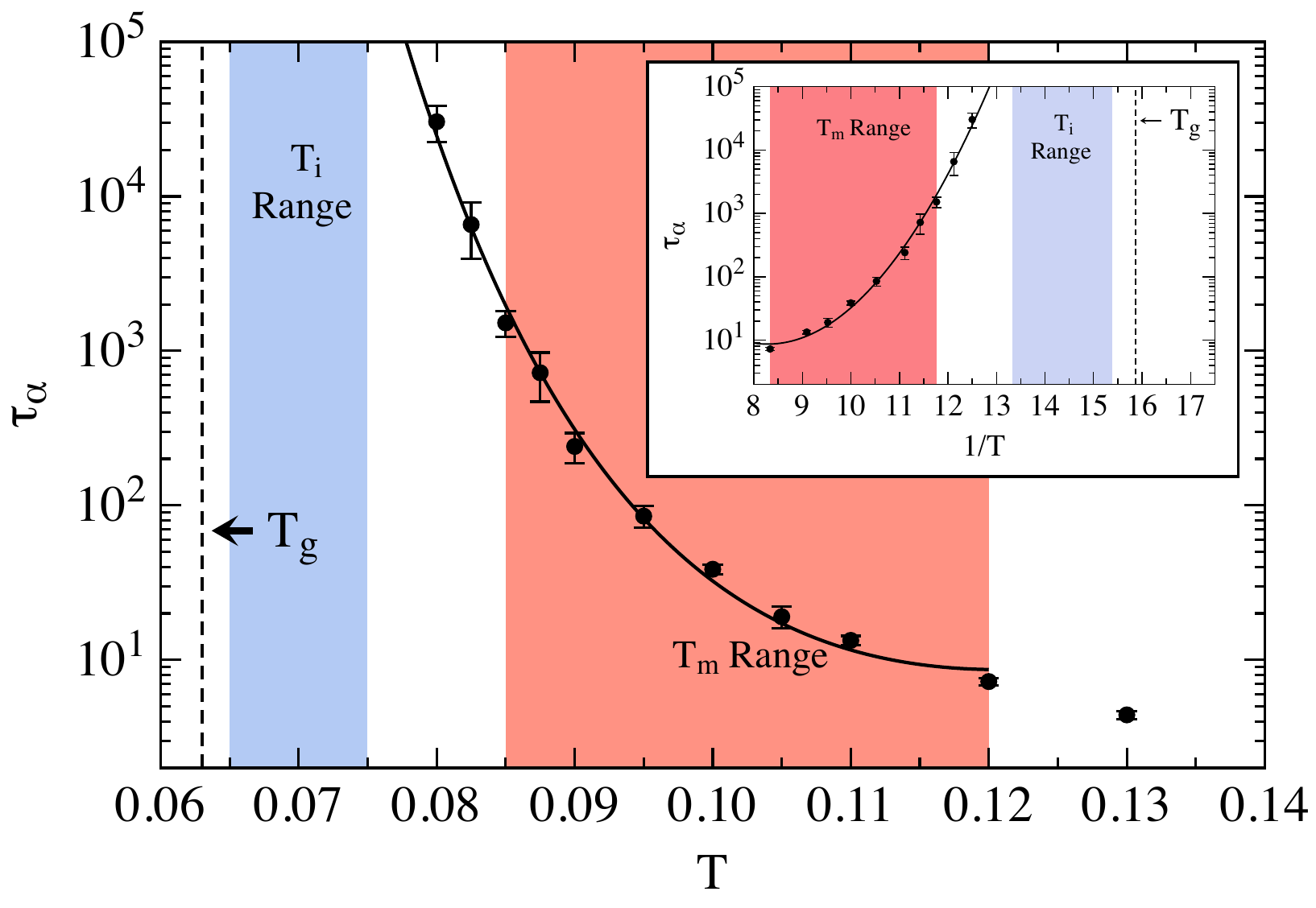}
\caption{\label{fig:tau} (Color online) Structural relaxation time, $\tau_\alpha$, for the equilibrated bulk system (circles) fitted with a parabolic law (solid line, see text). The estimated experimental $T_\mathrm{g}$ is shown with a dashed line. The blue box denotes the range of $T_\mathrm{i}$ over which bulk systems are prepared as film equilibrium precursors. The red box denotes the range of $T_\mathrm{m}$ over which films were melted. The inset presents the same information as a function of $1/T$ for reference.}
\end{figure}

In Fig.~\ref{fig:tau}, we use $\tau_\alpha(T)$ to determine the onset temperature by observing departure from high-temperature Arrhenius scaling, $\tau_\alpha = \tau_\infty e^{E_\infty/T}$ below $T_{\mathrm{on}} \approx 0.12$, with $E_\infty =0.57$. The experimental glass temperature $T_\mathrm{g}$ is not directly accessible in simulations, but can be reliably estimated by measuring dynamics over the accessible numerical window and fitting $\tau_\alpha$ to various functional forms~\cite{Ninarello2017}. Previous work has shown that the parabolic fit $\tau_\alpha = \tau_1 e^{E_p(1/T - 1/T_1)^2}$ yields an accurate extrapolation of $\tau_\alpha$ towards experimental timescales to obtain $T_\mathrm{g}$ as 
$\tau_\alpha(T_\mathrm{g})=10^{12}\tau_\alpha(T_{\mathrm{on}})$. We find $T_\mathrm{g}=0.063$, which sets a reference scale for film stability. 

To prepare films, we first run bulk simulations of systems using various thermal protocols at initial temperatures $T_\mathrm{i}$~\cite{SI}. These configurations are true equilibrium states for $T_\mathrm{i} \geq 0.065$, but only well-aged glasses for $T_\mathrm{i}=0.04$ where complete thermalization cannot be ensured. Films are then obtained by removing the periodic boundary condition in the $z$ direction, immobilizing a bottom layer of thickness $5$ to create a substrate, and leaving the top layer free. The resulting isotropic films have a height of about $42$ and the periodic box side in the orthogonal directions is about 40. 
After heating these systems from $T_\mathrm{i}$ to $T_\mathrm{m}$ at a rapid rate of $2 \cdot 10^{-4}$, they are finally held at $T_\mathrm{m}$ with an $NVT$ Nos\'e-Hoover thermostat until they melt. Because the melting 
temperatures lie below the onset of slow dynamics, a mildly supercooled liquids results. 

\begin{figure*}
\includegraphics[bb=0.5in 0.1in 9.9in 8.5in,clip=true,width=0.68\columnwidth]{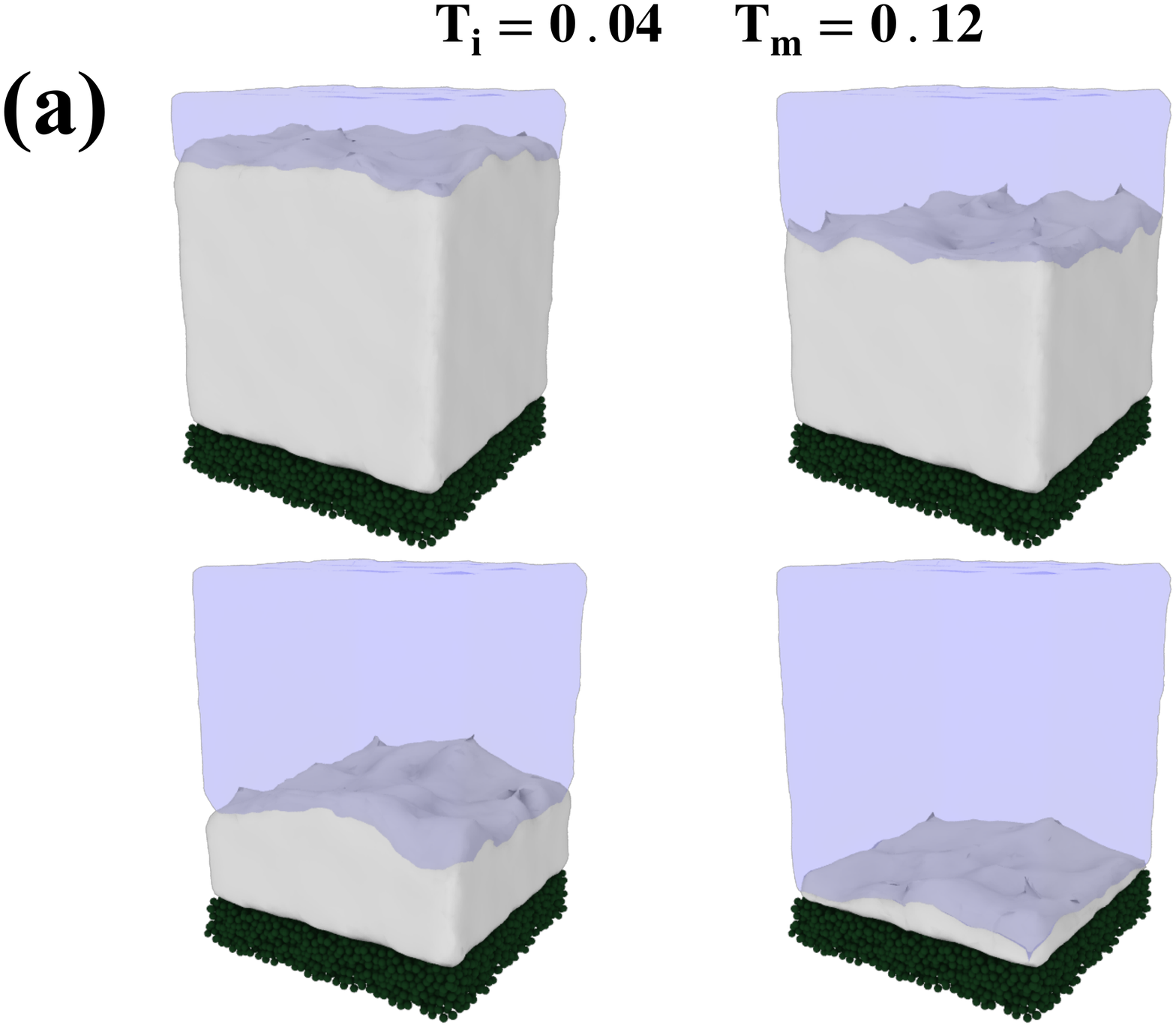} \vline
\includegraphics[bb=0.5in 0.1in 9.9in 8.5in,clip=true,width=0.68\columnwidth]{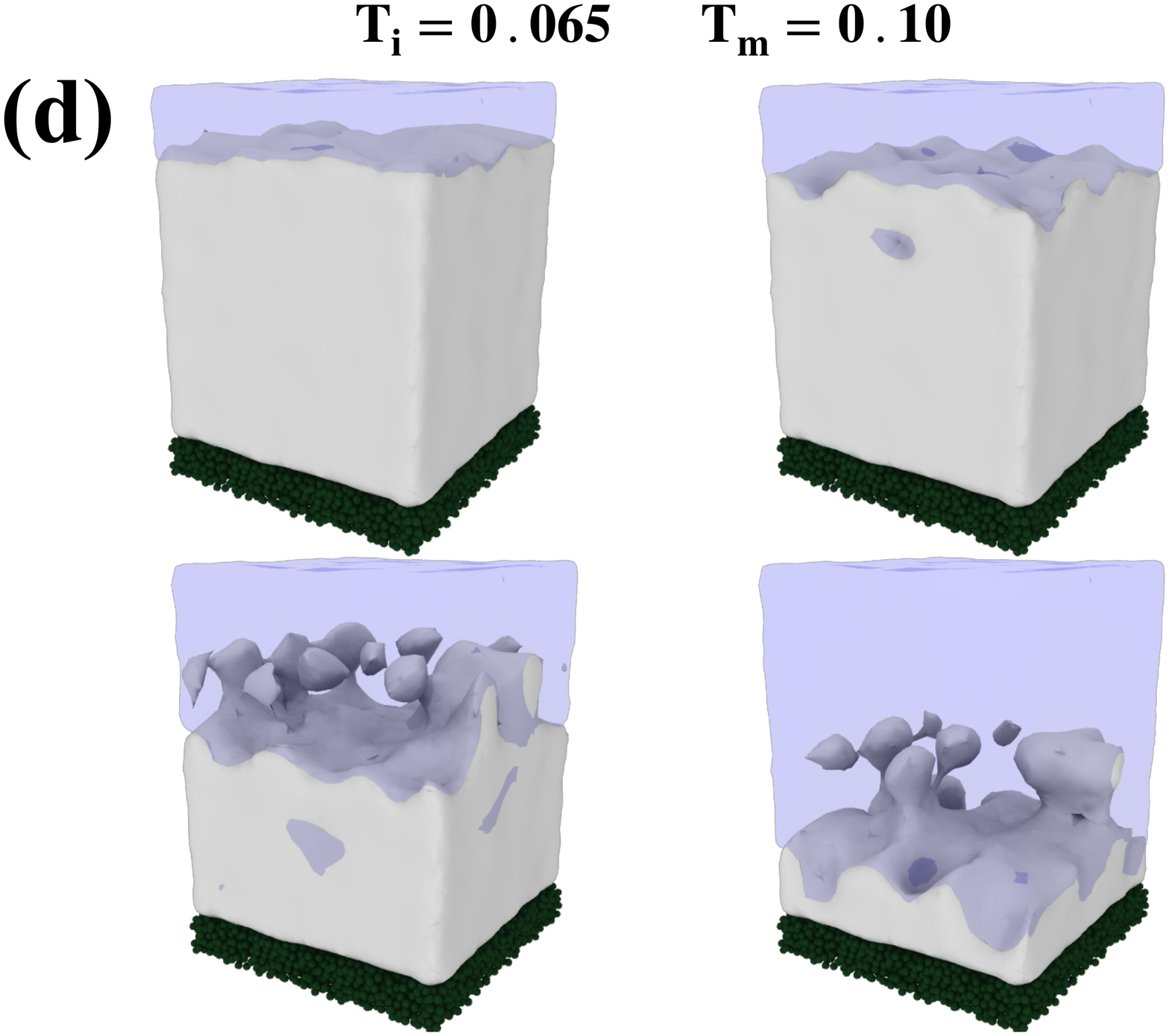} \vline
\includegraphics[bb=0.5in 0.1in 9.9in 8.5in,clip=true,width=0.68\columnwidth]{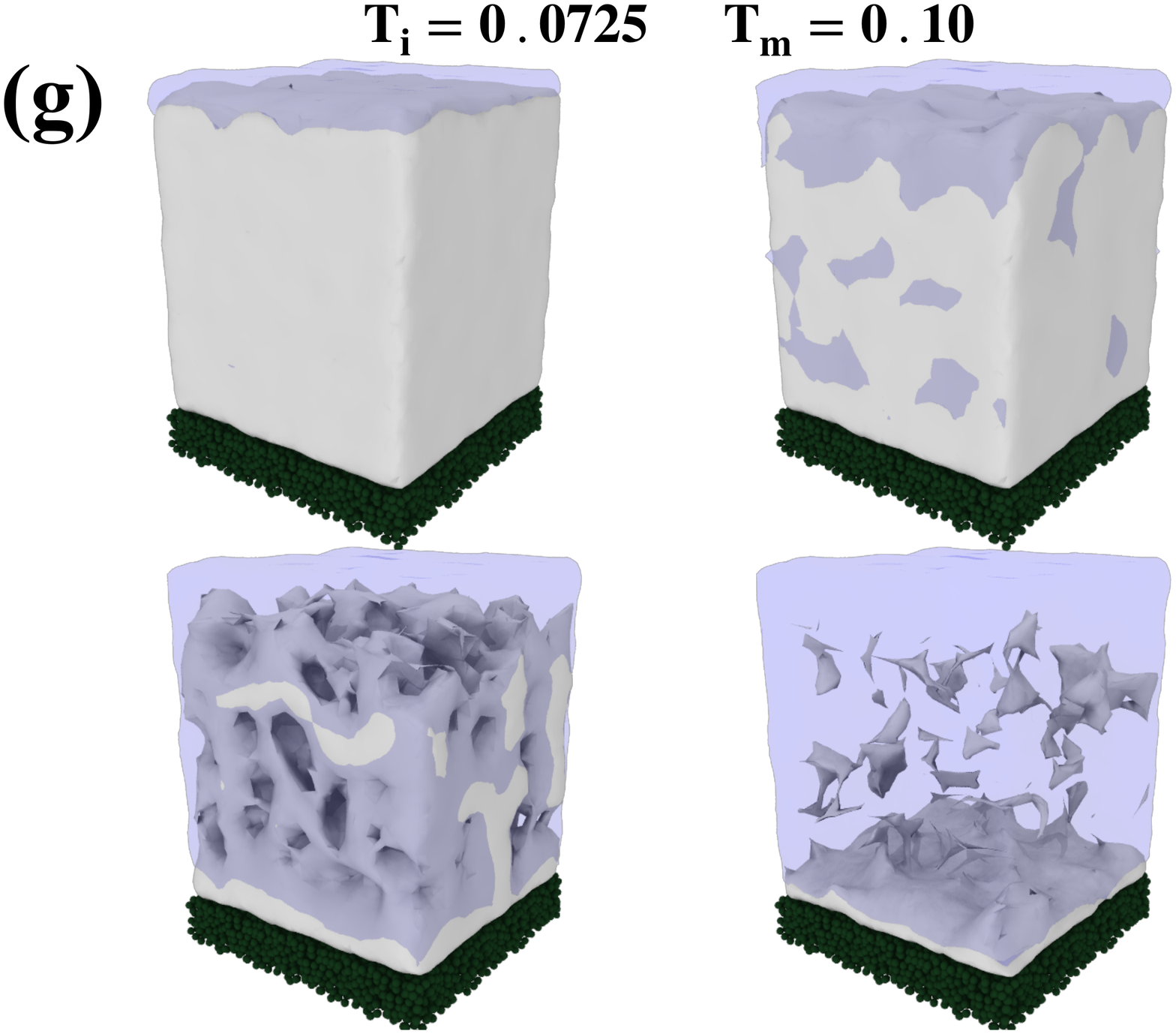} 
\includegraphics[width=0.68\columnwidth]{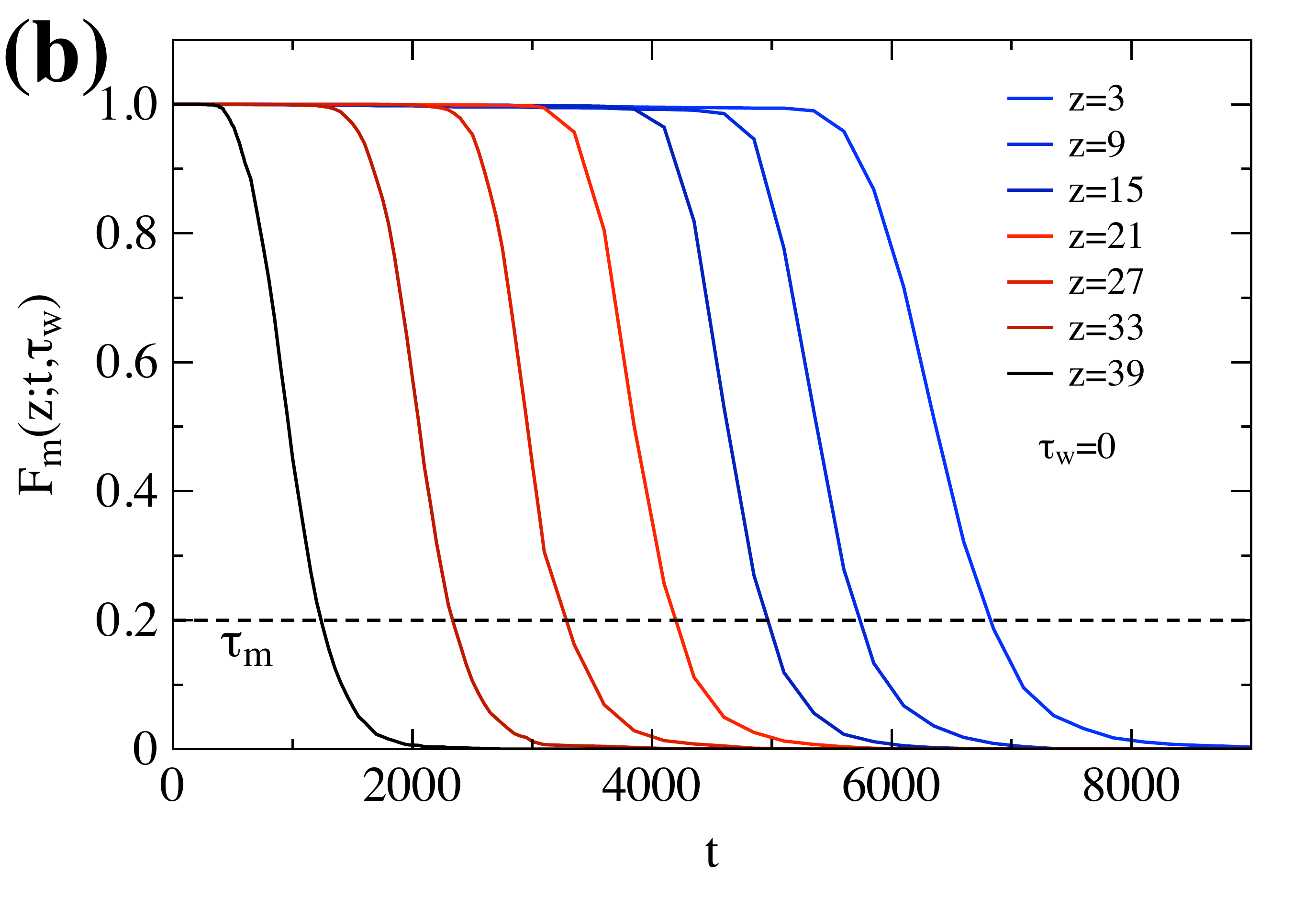} \vline
\includegraphics[width=0.68\columnwidth]{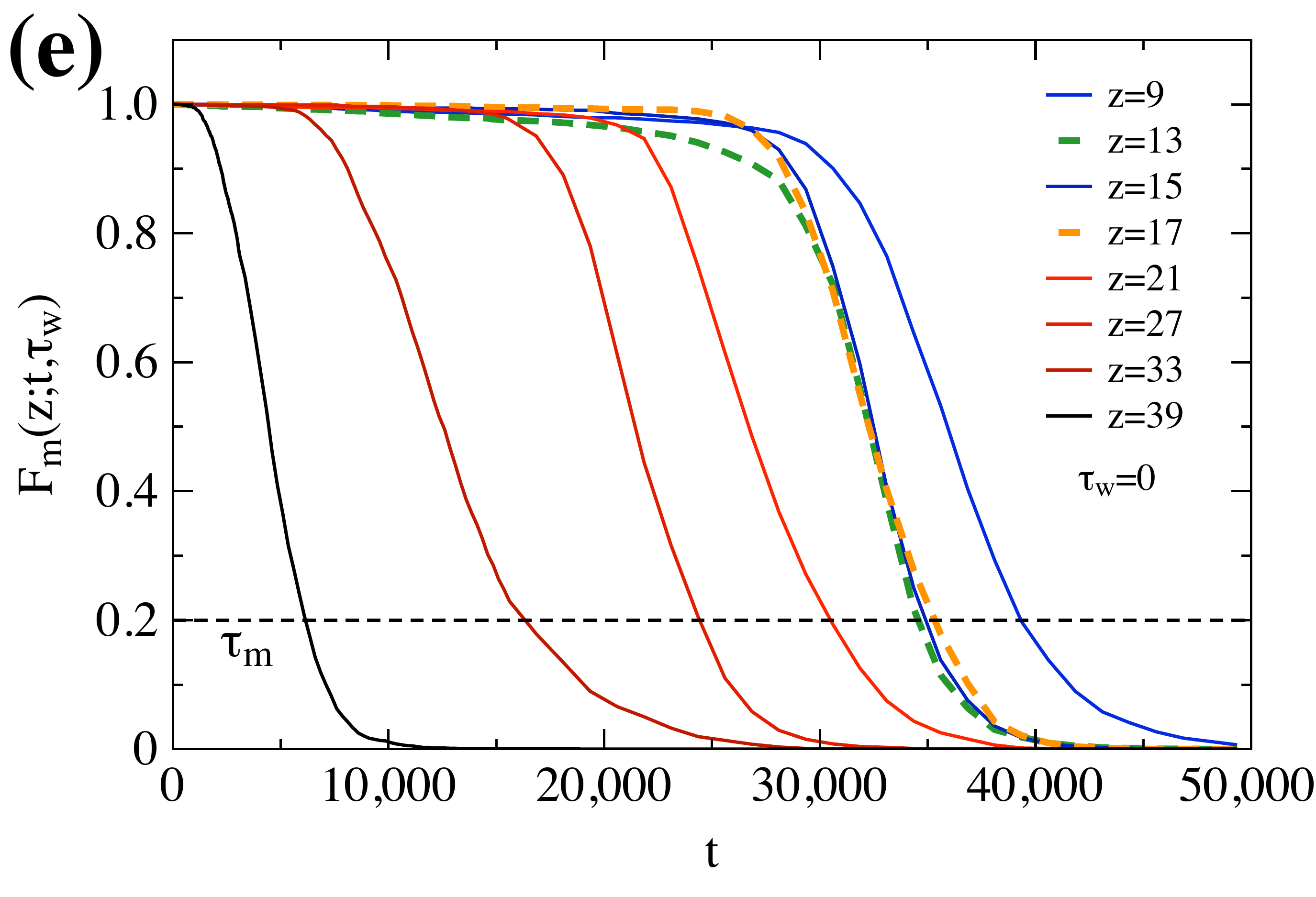} \vline
\includegraphics[width=0.68\columnwidth]{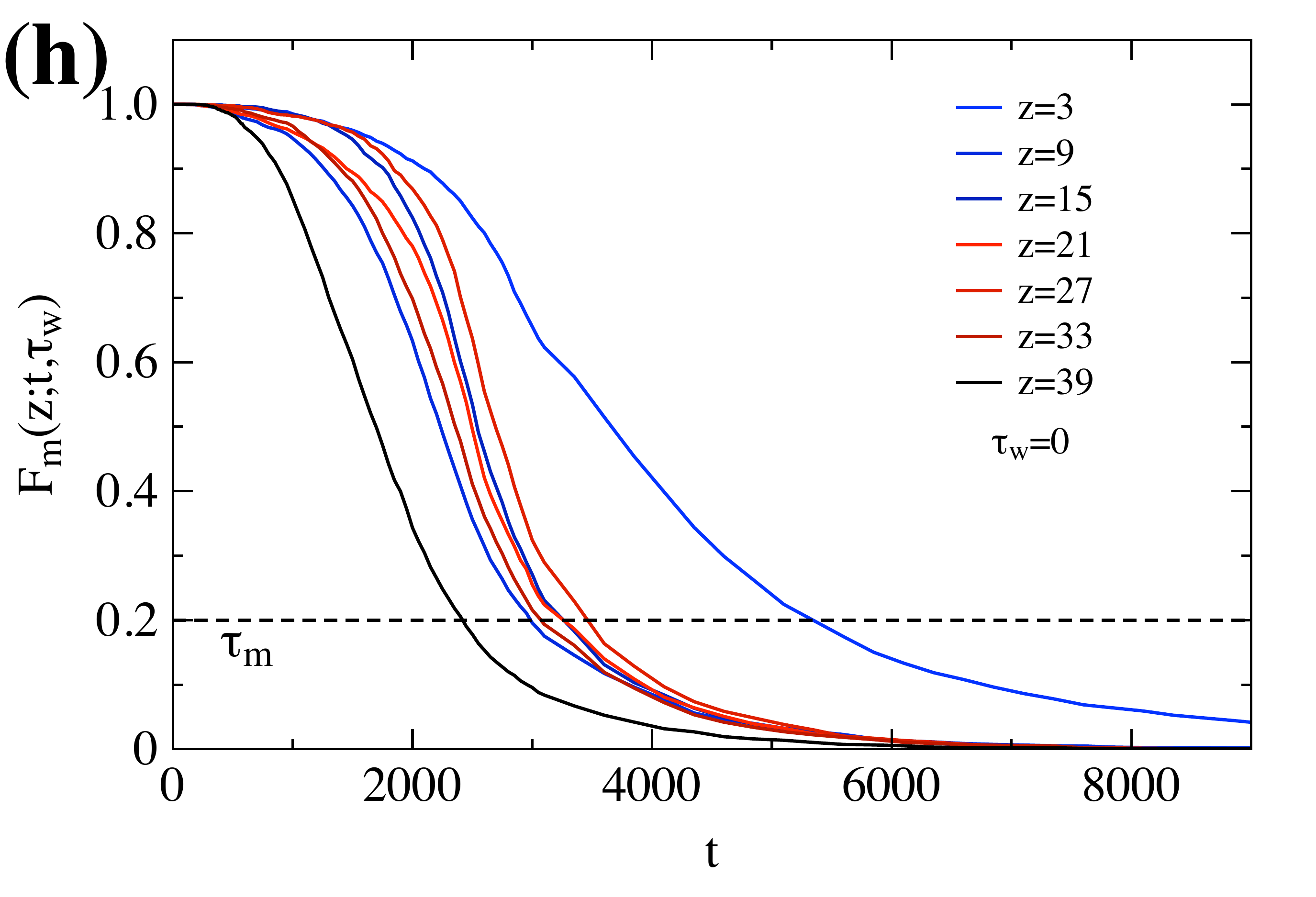}
\includegraphics[width=0.68\columnwidth]{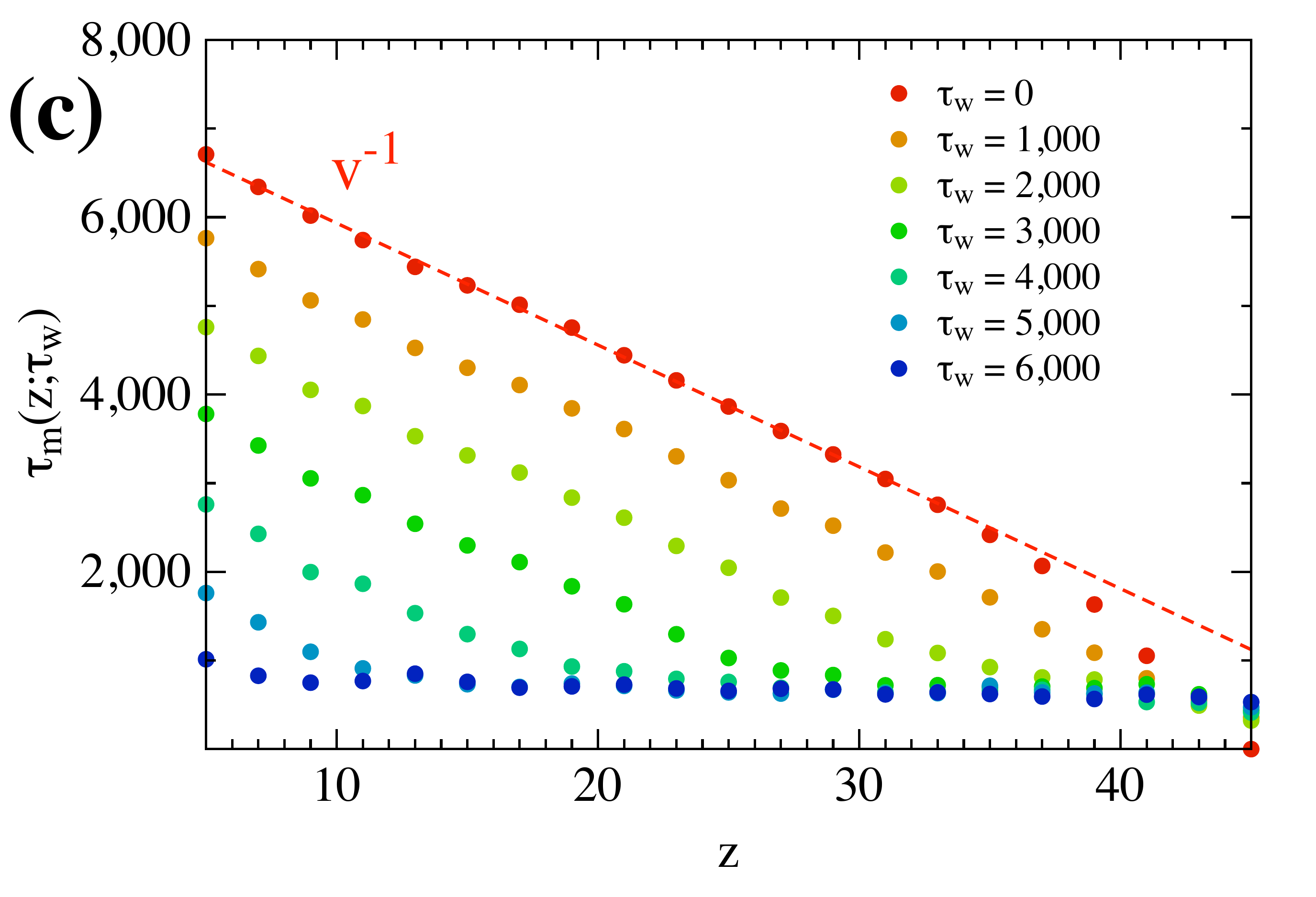} \vline
\includegraphics[width=0.68\columnwidth]{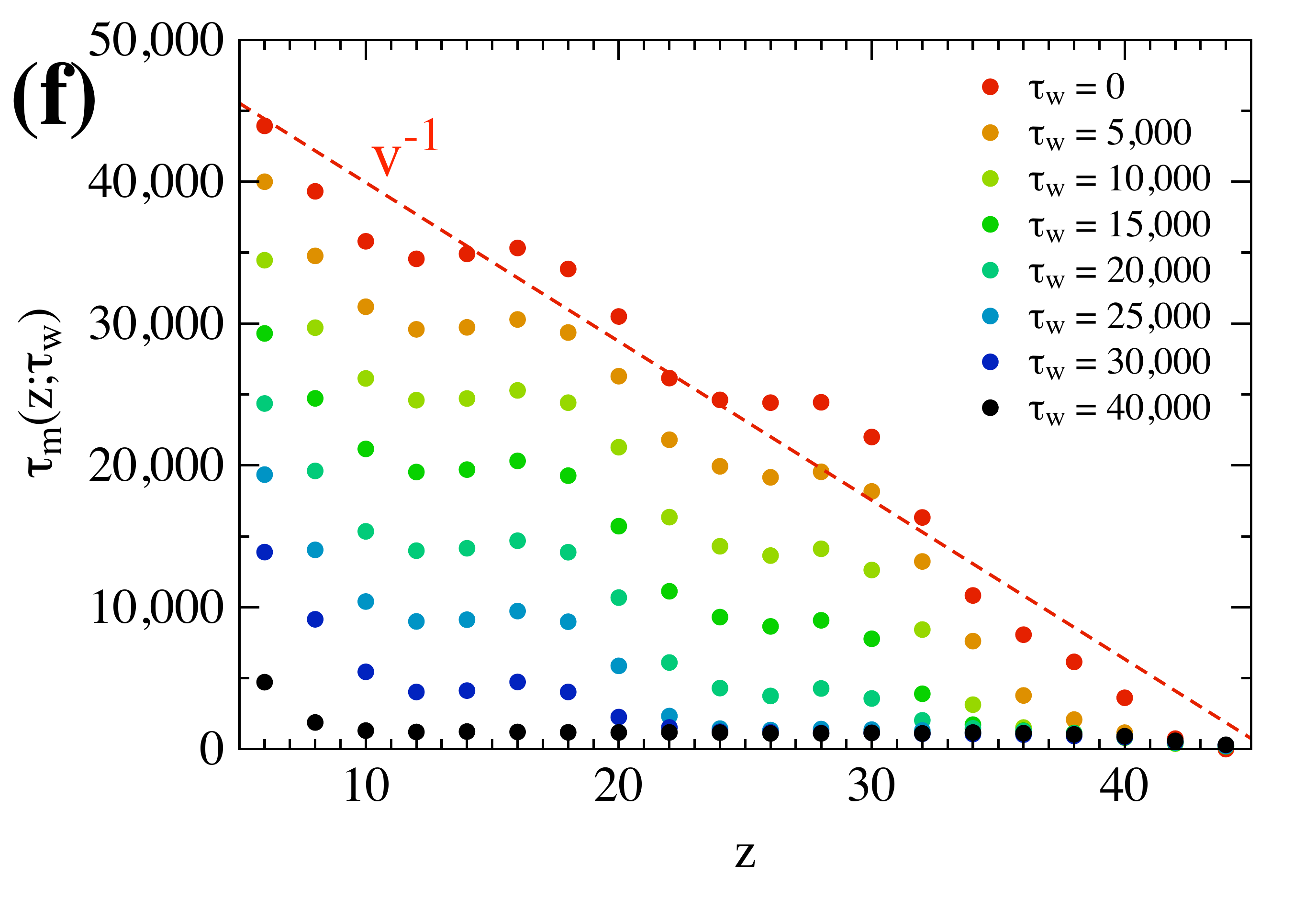} \vline
\includegraphics[width=0.68\columnwidth]{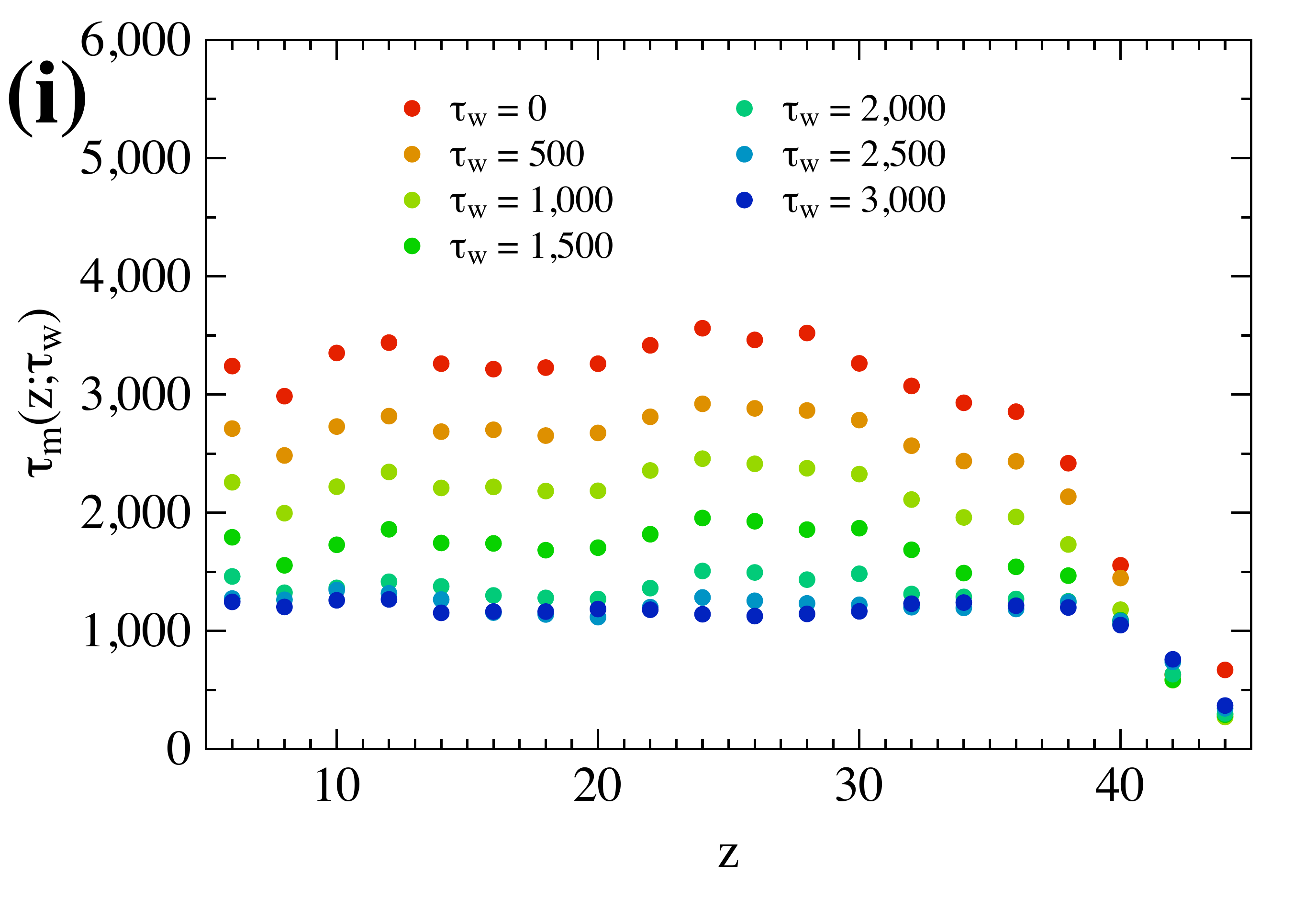}
\caption{\label{fig:tauw} (Color online)  Examples of glass film melting:
(Left) a pure melting front initiated at the surface; (Center) a front competing with bulk melting; (Right) pure bulk melting. 
(a,d,g) Representative snapshots of the three regimes are presented using transparent blue surfaces around regions of at least three mobile particles, opaque regions are glassy, and the dark green layer is the substrate.  
(b,e,h): overlap functions color coded based on the substrate distance: blue curves denote $z$ closest to the substrate and red curves those closest to the surface.
(c,f,i): melting times, $\tau_\mathrm{m}$, defined from the decay of the overlap function to 0.2 (dashed lines in middle row) for several waiting times, $t_\mathrm{w}$. A front propagation velocity (red dashed line) can be extracted in (c) and (f).}
\end{figure*}

The snapshots in Fig.~\ref{fig:tauw}(a,d,g) illustrate three melting regimes: 
(a) pure surface-initiated melting with a flat front that propagates over the entire film; 
(d) front-mediated melting competing with bulk melting;
(g) pure bulk melting with no apparent front. The numerical observation of these three regimes for films thicker than 40 particles is our central achievement. In particular, propagating fronts had only been
indirectly inferred in experiments, that lack the needed spatial resolution.  

To quantify these observations, we identify melted regions using a spatially-resolved overlap function:
\begin{eqnarray}
F_\mathrm{m}(z;t, \tau_\mathrm{w})  &= & 
 \frac{1}{N_z} \sum_{n=1}^{N_z} \prod_{\tau=\tau_\mathrm{w}}^t \Theta[a-\delta \rho_n(\tau)] \nonumber \\
 & \times & \Theta[z - (z_n(\tau_\mathrm{w})-1)]\Theta[(z_n(\tau_\mathrm{w})+1)-z], 
\nonumber 
\end{eqnarray}
where $\delta \rho_n(t) = \sqrt{[x_n(t)-x_n(\tau_\mathrm{w})]^2 + [y_n(t) - y_n(\tau_\mathrm{w})]^2}$, 
$\Theta$ is the Heaviside step function, and $N_z$ is the number of particles within a slab centered at $z$ of height $2$, after a waiting time $\tau_\mathrm{w}$. 
We choose $a = 1.6$, which is approximately equal to the largest particle diameter, to estimate the local structural relaxation. Note that if a particle moves more than $a$ in the horizontal direction, it is deemed mobile and no longer contributes to $F_\mathrm{m}(z;t,\tau_\mathrm{w})$, even if it eventually returns close to its original position. In Fig.~\ref{fig:tauw}(b,e,h) we show $F_\mathrm{m}(z;t,\tau_\mathrm{w}=0)$ for the three regimes. We define a relaxation time, $\tau_\mathrm{m}(z, \tau_\mathrm{w})$, as $F_\mathrm{m}(z;\tau_\mathrm{m},\tau_\mathrm{w}) =0.2$, and report its evolution in Fig.~\ref{fig:tauw}(c,f,i).  

Figure~\ref{fig:tauw}(b) illustrates that the functional decay of $F_\mathrm{m}(z;t,\tau_\mathrm{w}=0)$ is similar for all film depths in
systems with a sharp melting front. The associated decay time increases linearly with the distance from the film surface, see, e.g., Fig.~\ref{fig:tauw}(c). Deviations from a linear $z$ dependence only appear when $\tau_\mathrm{m}$ reaches $\tau_\alpha(T_\mathrm{m})$. Remarkably, the slope is independent of $\tau_\mathrm{w}$ and thus provides a robust estimate of the front velocity $v$. In practice, the front velocity is obtained by fitting $\tau_\mathrm{m} = \tau_\mathrm{m}^0 - z/v$, while making sure that the result is also consistent with the waiting time needed to fully melt the system.

By contrast, Fig.~\ref{fig:tauw}(e) does not show such a simple spatial dependence across the film. For instance, the decay of $F_\mathrm{m}(z;t,\tau_\mathrm{w}=0)$ at $z=13$-17 is essentially constant. The resulting profile is not perfectly linear, see, e.g., Fig.~\ref{fig:tauw}(f), but the decay times do decrease farther from the substrate, and hence a front velocity can still be extracted. This juxtaposition of behaviors is consistent with the emergence of a competing melting mechanism in the bulk.

In the opposite limit shown in Fig.~\ref{fig:tauw}(h), the top layer of the film melts rapidly and then, after a short time, the rest of the film essentially melts at once. Accordingly, the $z$ dependence of $\tau_\mathrm{m}(z,\tau_\mathrm{w})$ in Fig.~\ref{fig:tauw}(i) confirms that the decay time is constant throughout the film. Melting is then a purely bulk process.  

Evidence of front melting was obtained over a broad range of $(T_\mathrm{i}, T_\mathrm{m})$. For a given $T_\mathrm{i}$, a crossover from heterogeneous to homogeneous melting occurs as $T_\mathrm{m}$ increases, over a range $\Delta T_\mathrm{m} \approx 0.005$. Systems in this crossover region exhibit a melting front, but this front only relaxes a finite fraction of the film. It is however difficult to directly characterize $\ell_\mathrm{c}$, because the front is no longer sharply defined when the bulk process takes over. 

\begin{figure}
\includegraphics[width=1.\columnwidth]{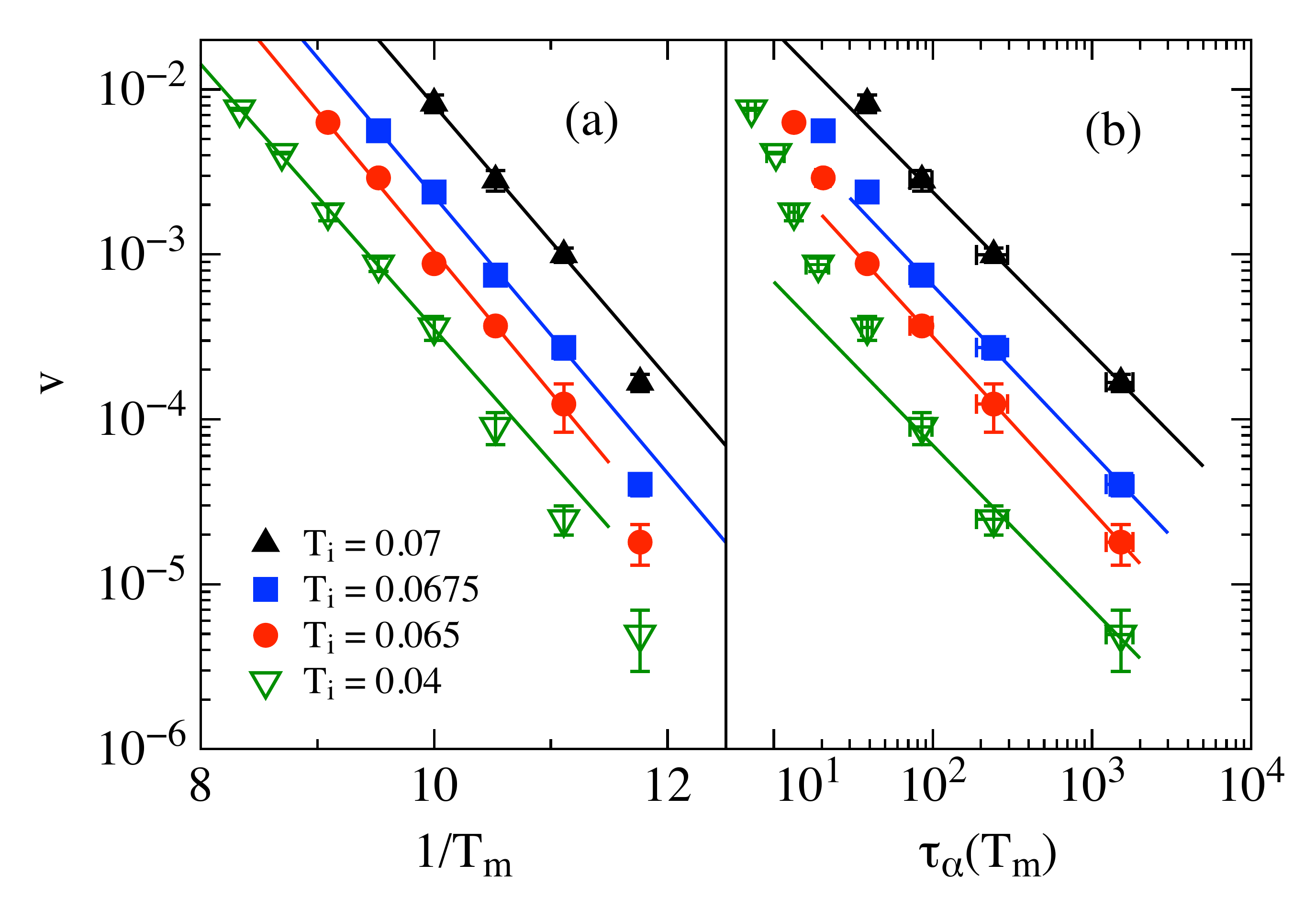}
\caption{\label{velocity} (Color online) Melting front velocity versus (a) $1/T_\mathrm{m}$ and (b) $\tau_\alpha(T_\mathrm{m})$ with, respectively, 
Arrhenius fits, $v = v(T_\mathrm{i})e^{-E_\mathrm{a}/T_\mathrm{m}}$, with $E_\mathrm{a} \approx 2.0$ for all $T_\mathrm{i}$, and power-law fits, $v = v_0(T_\mathrm{i}) \tau_\alpha^{-\gamma}$, with $\gamma=1$ for all $T_\mathrm{i}$.}
\end{figure}

The evolution of $v(T_\mathrm{i},T_\mathrm{m})$ is reported in Fig.~\ref{velocity}. 
If the melting temperature is large enough, $T_\mathrm{m} > 0.095$, an Arrhenius description with $E_\mathrm{a}=2.0 \pm 0.1$ captures our data well for all $T_\mathrm{i}$ (Fig.~\ref{velocity}(a)). This activation energy is about four times higher than that inferred from the temperature dependence of the relaxation time above $T_{\mathrm{on}}$, which suggests that the energy barriers overcome during melting are different from those of the bulk relaxation process at the same temperature. In addition, as found in experiments the kinetic stability of the glass only enters as a prefactor to the Arrhenius scaling, with more stable systems exhibiting slower front propagation. 
The Arrhenius scaling breaks down for $T_\mathrm{m} < 0.095$ for all $T_\mathrm{i}$. Following experimental observations, we then fit $v$ to Eq.~\eqref{eq:alpha} and find $\gamma \approx 1$ (Fig.~\ref{velocity}(b)), which is compatible with the only available numerical data~\cite{Hocky2014}. The accessible range of $v$ is, however, too small to fully validate the power-law scaling and obtain an accurate estimate of $\gamma$. These results nonetheless indicate that the front velocity is controlled by the supercooled liquid dynamics for $T \lesssim 0.095$. Here again, the glass stability only enters as a prefactor. 

Whereas more stable glasses entirely melt with a moving front, the snapshots in Fig.~\ref{fig:tauw} (d,g) reveal that large melted domains can appear ahead of the front. This directly shows that even without a free surface, a bulk process eventually melts the films in a finite time, $\tau_\mathrm{m}^{\mathrm{bulk}}$. The length scale $\ell_\mathrm{c}$ that characterizes the crossover between front propagation and bulk melting is then given by $\ell_\mathrm{c} = v \tau_\mathrm{m}^{\mathrm{bulk}}$~\cite{Hocky2014,Jack2016}. In order to estimate $\ell_\mathrm{c}$ we independently determine $\tau_\mathrm{m}^{\mathrm{bulk}}$ using bulk simulations, and combine the results with the above measurements of $v$~\cite{SI}. We do not find any evidence of a finite size effect in $\tau_\mathrm{m}^{\mathrm{bulk}}$ for
$T_i = 0.065$ and 0.0675, and finite-size effects cannot be disentangled from aging for $\tau_\mathrm{m}^{\mathrm{bulk}}$ for $T_i = 0.04$. 

\begin{figure}
\includegraphics[width=1.\columnwidth]{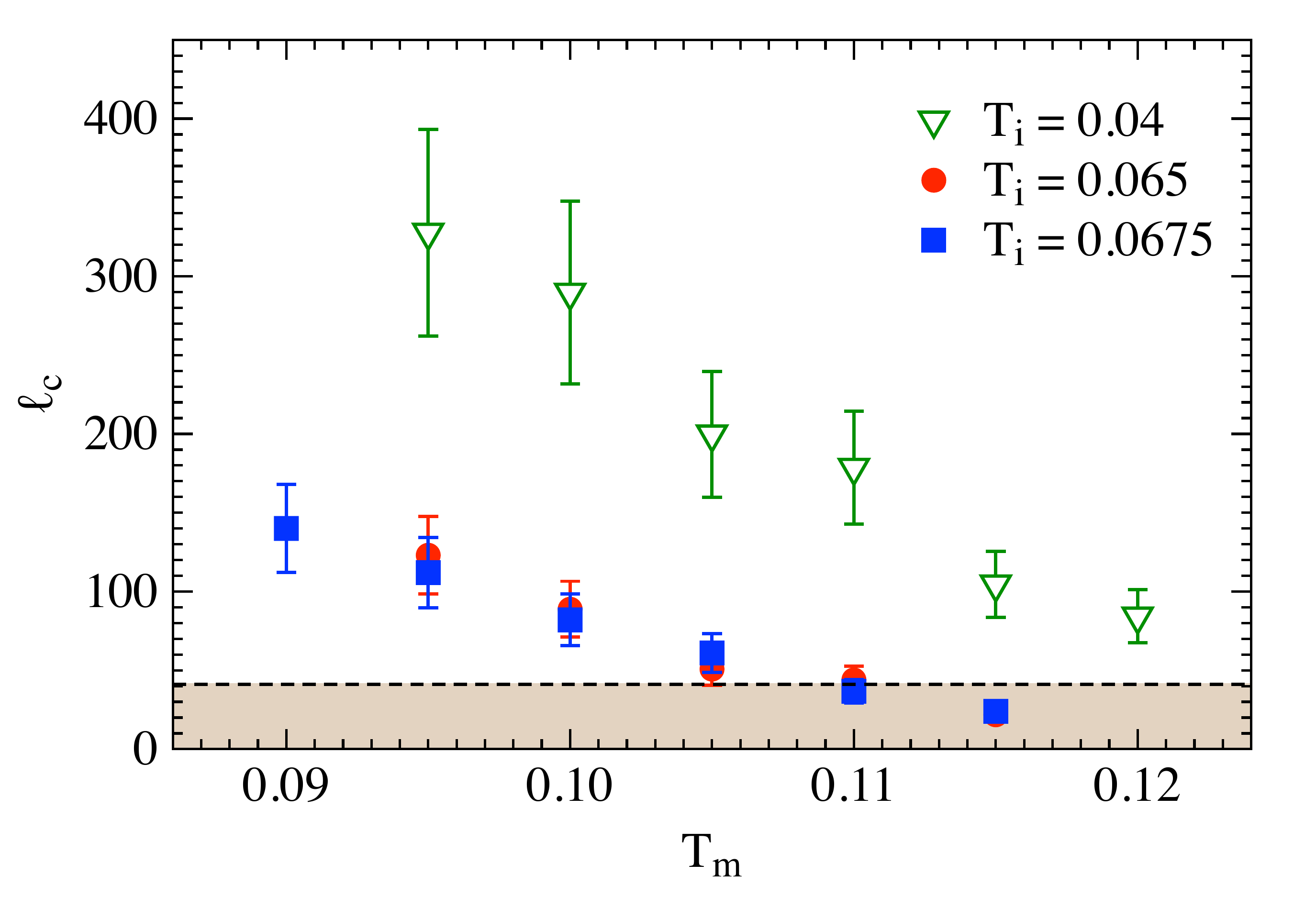}
\caption{\label{fig:length}
(Color online) The crossover length $\ell_\mathrm{c}$ characterizes the maximum film thickness for where front melting dominates. The brown region denotes the range of $\ell_\mathrm{c}$ smaller than the film thickness considered here. In this regime, front-mediated melting competes with bulk melting.}
\end{figure}
 
In Fig.~\ref{fig:length}, we report that $\ell_\mathrm{c}$ grows both upon increasing glass stability (lower $T_\mathrm{i}$) and upon lowering $T_\mathrm{m}$. While the kinetics of both front-mediated and bulk melting slow down with decreasing $T_\mathrm{m}$ and $T_\mathrm{i}$, the fact that $\ell_\mathrm{c}$ grows indicates that bulk melting is more strongly suppressed than front melting. In bulk melting, regions of liquid form and grow within the glass, meaning that $\tau_\mathrm{m}^{\mathrm{bulk}}$ is determined by a non-trivial combination of the two timescales associated with the formation and the growth of these regions~\cite{Jack2016,Fullerton2017}. Too little is known about how $T_\mathrm{i}$ and $T_\mathrm{m}$ affect these two timescales individually to untangle their influence. 

To compare our observations with experiments, we consider toluene films~\cite{Dalal2013,Dalal2012,Dawson2011}. 
For toluene, we take $\sigma_0 \approx 0.6$nm \cite{Jorgensen1992}, R\'afols-Rib\'e \textit{et al.}~\cite{Rafols2017} report stability-dependent $\ell_\mathrm{c}$ in the range 50-200nm, and Bhattacharya \textit{et al.}~\cite{Bhattacharya2014} find $\ell_\mathrm{c}\approx 250$nm. The largest length scale we measure for $T_\mathrm{i}=0.04$ and $T_\mathrm{m} = 0.095$ gives $\ell_\mathrm{c} = 375 \sigma_0 \approx 225$nm, which compares favorably with the measurements made
using nanocalorimetry for samples vapor deposited around $0.95T_\mathrm{g}$ \cite{Rafols2017,marta-new}. The result of 375 particle diameters is 
in the middle of the range of 20 to 2000 particle diameters reported in experiments \cite{Sepulveda2014,Sepulveda2012}. We rewrite $\ell_\mathrm{c} = v \tau_{\rm m}^{\rm bulk} = \ell_o {\cal S}$~\cite{Hocky2014}, which is the product of a length $\ell_o = v \tau_\alpha$ and the stability ratio ${\cal S} = \tau_{\rm m}^{\rm bulk}/\tau_\alpha$. We compiled experimental and numerical data from $T_{\rm on}$ to $T_g$, and found that $\ell_o$ is a microscopic length with weak temperature dependence (it is controlled by the small exponent $1-\gamma$). From Fig.~\ref{velocity}, we estimate $\ell_o \approx 0.01-0.1$, which is again comparable to experiments~\cite{Rodriguez2015,Rodriguez2014,Tylinski2015}. 
As a result, the crossover length $\ell_\mathrm{c}$ is mainly controlled by the stability ratio. Since ${\cal S}$ may increase up to $10^{5}$~\cite{Fullerton2017,Sepulveda2014}, the observation of very large $\ell_\mathrm{c}$ therefore directly reflects kinetic ultrastability~\cite{marta-new}. Although we cannot perform simulations at lower $T_{\rm m}$, our simulations are therefore in qualitative agreement with the very large crossover length estimated for substances such as methyl-$m$-toluate and indomethacin.


A recent algorithmic progress~\cite{Ninarello2017} allowed us to create films that are sufficiently large and stable to observe front-mediated and bulk melting, as well as the competition between them. We have thus replicated, with atomic resolution, a salient experimental feature of vapor-deposited glasses~\cite{Swallen2009,Kearns2010}. Because our samples are obtained from isotropic, equilibrium bulk simulations, we conclude that front-mediated melting can result from kinetic stability alone, independently of the vapor-deposition process and of the non-equilibrium or anisotropic nature of the glass. It should thus also be observable in conventional glasses at properly chosen melting temperatures, contradicting the hypothesis (stemming from lack of spatial resolution in experiments) that liquid-cooled glasses do not melt via a front. We have further characterized the maximum film thickness that melts via a surface-initiated front before bulk melting becomes competitive. 
As observed in experiments, an increase in $\ell_\mathrm{c}$ is directly linked to the growing kinetic stability of ultrastable glasses. Our results show that a detailed characterization of the propagating front properties are now possible, and suggest also that studies of bulk melting, which has received much less experimental attention, could provide further insights. 


\acknowledgments

We thank M. Ediger and J. Rodr\'iguez-Viejo for useful exchanges about experiments. We acknowledge support of NSF Grant No.~DMR-1608086 (EF) and the Simons Foundation (\# 454933, L. B., and \# 454937, P. C.). This research utilized the CSU ISTeC Cray HPC System supported by NSF Grant No.~CNS-0923386. 
Data associated with this work are available from the Duke Digital Repository \cite{ddr}.

\end{document}